\documentclass[twocolumn]{emulateapj}
\usepackage{epsfig,psfig,amsmath,amsfonts,amssymb}
\newcommand{\com}[1] {#1}
\newcommand{\simgt}{\,\hbox{\lower0.6ex\hbox{$\sim$}\llap{\raise0.6ex\hbox{$>$}}}\,}
\newcommand{\simlt}{\,\hbox{\lower0.6ex\hbox{$\sim$}\llap{\raise0.6ex\hbox{$<$}}}\,}
\begin{document}
\shorttitle{Weak Lensing: Ground {\it vs} Space}
\title{A Comparison of Weak Lensing Measurements\\From Ground- and Space-Based Facilities}
\shortauthors{M.\ M.\ Kasliwal et al.}
\author{Mansi M.\ Kasliwal\altaffilmark{1,2}, 
        Richard Massey\altaffilmark{1}, 
        Richard S.\ Ellis\altaffilmark{1},
        Satoshi Miyazaki\altaffilmark{3} \& 
        Jason Rhodes\altaffilmark{4,1}}

\altaffiltext{1}{Astronomy Department, California Institute of Technology, 105-24, Pasadena, CA 91125, 
USA}
\altaffiltext{2}{George Ellory Hale Fellow of Moore Foundation}
\altaffiltext{3}{National Astronomical Observatory of Japan, Mitaka, Tokyo 181-8588, Japan}
\altaffiltext{4}{Jet Propulsion Laboratory, California Institute of Technology, 105-24, Pasadena, CA 
91125, USA}


\begin{abstract}

We assess the relative merits of weak lensing surveys, using overlapping imaging data 
from the ground-based Subaru telescope and the Hubble Space Telescope (HST). 
Our tests complement similar studies undertaken with simulated data. 
From observations of 230,000 matched objects in the 2 square degree COSMOS field, 
we identify the limit at which faint galaxy shapes can be reliably measured 
from the ground. Our ground-based shear catalog achieves sub-percent calibration bias compared to 
high resolution space-based data, for galaxies brighter than $i^{\prime}\simeq$24.5 and with 
half-light radii larger than $1.8\arcsec$. This selection corresponds to a surface density of 
15 galaxies arcmin$^{-2}$ compared to $\sim 71$ arcmin$^{-2}$ from space. On the other hand
the survey speed of current ground-based facilities is much faster than that  of HST, although 
this gain is mitigated by the increased depth of space-based imaging desirable for tomographic
(3D) analyses. As an independent experiment, we also reconstruct the projected mass distribution
in the COSMOS field using both data sets, and compare the derived cluster catalogs with those 
from $X$-ray observations. The ground-based catalog achieves a reasonable degree of completeness, 
with minimal contamination and no detected bias, for massive clusters at redshifts $0.2<z<0.5$. 
The space-based data provide improved precision and a greater sensitivity to clusters of lower 
mass or at higher redshift.
\end{abstract}



\section{Introduction}
\label{sec:Introduction}

Dark matter dominates the gravitational component of the cosmic energy density and thus provides
the framework for structure formation in the Universe. However, by its nature, the distribution
and cosmic growth are challenging to observe. The most promising probe is weak gravitational
lensing: analysis of the distorted shapes of ordinary galaxies behind foreground mass
concentrations. Several numerical techniques are now available to recover the projected mass
distribution from these distortions, and tests on simulated datasets are underway to verify their
precision \citep{step1,step2}. There is great optimism in the weak lensing community that such
methods will enable both the tomographic mapping of dark matter structures in time and space. This
will also provide a robust statistical measure of the nature of dark energy over redshifts 0$<z<$1
\citep{mellier99, refregier03}

Observational progress has been particularly dramatic. The first detections of 
statistical ``cosmic shear'' were only published in 2000 \citep{bacon00, kaiser00, wittman00,
vanwaerbeke00}. In the subsequent \com{seven} years, weak lensing surveys have measured the dark 
matter power spectrum \citep{brown03, heymans05, hoekstra06, sembolini06}, traced 
the evolution of structure \citep{bacon05, kitching06, massey07a}, enabled the construction 
of lensing-selected cluster catalogs \citep{miyazaki02a, wittman06, schirmer07, miyazaki07}, 
and non-parametrically reconstructed the total mass distribution both in clusters 
\citep{kneib03, clowe06, jee07} and on larger scales \citep{massey07b}. As a result, 
weak lensing has been identified as the most promising route to understanding the nature 
of dark energy by the ESA-ESO Working Group on Fundamental 
Cosmology\footnote{\tt http://www.stecf.org/coordination/esa\_eso/cosmology.php},
joint NSF-NASA-DOE Astronomy and Astrophysics Advisory 
Committee\footnote{\tt http://www.nsf.gov/mps/ast/aaac.jsp}, 
and NSF-DOE High Energy Physics Advisory Panel 
Dark Energy Task Force\footnote{\tt http://www.nsf.gov/mps/ast/detf.jsp}. 

The primary signal of any weak lensing analysis is the statistically coherent 
distortion of background galaxies along adjacent lines of sight. The main sources of statistical 
noise are the finite density of galaxies that can be sufficiently well-detected and resolved for 
accurate shape measurement, plus their intrinsic morphologies. The density of resolved 
galaxies also governs the angular resolution and fidelity of a reconstructed mass map
which, in turn, determines the limiting halo mass that can be detected. On the other hand, 
statistical analyses of the dark matter power spectrum are less concerned with individual halos 
but require panoramic fields to counter the effects of cosmic (sample) variance. Minimizing 
statistical errors in such an analysis, within a finite survey lifetime, requires an optimal balance 
between area and depth.

A key debate in the development of future weak lensing experiments concerns the relative merits of
ground- versus space-based platforms. 
Ambitious surveys now being planned with dedicated, ground-based facilities (eg VST-KIDS, DES, 
Pan-STARRS, LSST). These are driven by technological progress including panoramic cameras with
small  optical distortions, highly sensitive imaging detectors, and (in the case of Pan-STARRS) 
on-chip active correction to reduce the width of the point spread function (PSF). Future surveys 
spanning significant fractions of the celestial sphere are envisaged, promising tight constraints 
on the  cosmological parameters.

However, measurements with current ground-based facilities are limited by the size and temporal
variations of the PSF. There is concern in many quarters that wide-field facilities operating in
space (e.g.\ DUNE, SNAP, JDEM) will ultimately be required to achieve the precision required
(particularly) to distinguish between various models of dark energy. Space-based facilities will
be more costly but will likely offer increased depth, better photometric performance and a stable
PSF. The key issue in gauging their merits is not statistical error, but the extent to which
potential biases in ground-based data may act as a ``systematic floor'' to prevent complete
exploitation.

Some valuable answers can be obtained by comparing simulated ground and space-based images, 
\citep{wittman05,lampton06} and the Shear TEsting Programme \citep[STEP:][]{step1,step2}.
However, the input parameters used to generate the simulated data may not be realistic or address
all the instrumental idiosyncrasies. Of particular concern are the stability and vagaries of the
PSF. No simulations have  yet adequately addressed this point -- which may, ultimately, be 
the limiting problem for ground-based data. It is often argued that future facilities will be 
carefully designed to mitigate any limitations realized with current observational facilities. 
While progress can no doubt be expected, both on the ground and in space, we believe many lessons
 can be learned from extant data and hardware with proven engineering pedigree.

In this paper, we present the first direct comparison of weak lensing analysis  {\it for the same
sky field} using ground and space-based data. Deep, panoramic imaging has been obtained for the
1.64 deg$^2$ COSMOS field \citep{scoville07a} by both the Advanced Camera for  Surveys ({\it ACS})
on board the Hubble Space Telescope (HST) \citep{scoville07b} and the {\it Suprime-Cam} imager at
the prime focus of the Subaru 8.2m telescope \citep{taniguchi07}. In both cases, the 
entire field was covered by mosaicing many
independent exposures.  The SuPrimeCam instrument was constructed with weak lensing analysis
particularly in mind, and currently provides the best image performance available from any
ground-based telescope, in terms of optical distortions over a large field. A comparison of these
datasets should therefore provide a realistic and valuable assessment of the relative performance
of state-of-the-art imagers on the ground and in space.

The paper is organized as follows. In \S\ref{sec:theory}, we briefly review the relevant theory. In
\S\ref{sec:data}, we describe the two data sets, data reduction pipelines and weak lensing
analyses. We then present the results. In \S\ref{sec:analyses}, we compare shear measures on a
galaxy-by-galaxy  basis to determine the optimum depth at which the ground-based data matches the 
performance of the (deeper) space  based data. This permits us to determine the relative survey
speeds of Subaru and HST for high precision cosmic shear experiments. In \S\ref{sec:Maps}, we 
construct maps of the mass distribution, treating the Subaru and HST maps as independent probes of
the same field, and contrast these against X-ray data. This permits us to evaluate the completeness
and reliability of a lensing-selected halo catalog, and evaluate the precision of their inferred
masses as a function of redshift. In \S\ref{sec:conc}, we summarize our results and discuss their
wider implications for future missions.

\section{Review of Weak Lensing Theory}\label{sec:theory}

Gravitational lensing by foreground mass structures distorts an image plane of distant galaxies
$I(\mathbf{x})$ via a coordinate transformation
\begin{equation}
\label{eqn:psi_def_theory}
{\mathcal A}_{ij} = \delta_{ij} + \frac{\partial (\delta x_i)}{\partial x_j}
  = \left( \begin{array}{cc}
1 - \kappa - \gamma_{1} & \gamma_{2}\\
\gamma_{2} & 1 - \kappa + \gamma_{1} \\
\end{array} \right),
\end{equation}
\noindent where $\delta x_i(\mathbf{x})$ is the deflection angle of the light rays. The {\em
convergence}
\begin{equation}
\label{eqn:kappa}
\kappa(\mathbf{x})=\frac{4\pi G}{2c^2}\int g(z) \rho(\mathbf{x},z) {\rm d}z ~,
\end{equation}
\noindent describes overall dilations and contractions. It is proportional to the 
total mass density $\rho$ projected along a line of sight, where the {\em lensing sensitivity function}
\begin{equation}
g(z)=\frac{2D_LD_{LS}}{D_S}
\label{eqn:sensitivitydefn}
\end{equation}
\noindent reflects the efficiency of foreground gravitational lenses at different redshifts --
containing a ratio of the angular diameter distance to a lens, the background source, and between the 
two. 
This can be more simply written as
\begin{equation}
\kappa ~ \equiv ~ \frac{1}{2} \left( \frac{\partial^2\Psi}{\partial x^2} + \frac{\partial^2\Psi}{\partial y^2}\right) 
~,
\label{eqn:convergencedefn}
\end{equation}
\noindent in terms of a 2D, projected version $\Psi(\mathbf{x})$ of the Newtonian gravitational potential.
Two components of {\em shear}
\begin{equation}
\big\{\gamma_1,~\gamma_2\big\}
 \equiv \Bigg\{ 
 \frac{1}{2} \left(\frac{\partial^2\Psi}{\partial x^2} - \frac{\partial^2\Psi}{\partial y^2}\right),~
        \frac{\partial^2\Psi}{\partial x\partial y}
        \Bigg\}~,
\label{eqn:sheardefn}
\end{equation}
\noindent describe stretches and compressions along (at $45^{\circ}$ from) the $x$-axis.

The observed shapes of background galaxies can be described by combination of their
Gaussian-weighted quadrupole moments
\begin{equation}
\label{eqn:ksbd}
d\equiv\frac{\iint I({\mathbf x}) ~W({\mathbf x})~r^2~{\mathrm d}^2{\mathbf x}}
             {\iint I({\mathbf x}) ~W({\mathbf x})~{\mathrm d}^2{\mathbf x}} ~,
\end{equation}
\begin{equation}
\label{eqn:ksbe}
\big\{\varepsilon_1,~\varepsilon_2\big\}
  \equiv\frac{\iint I({\mathbf x}) ~W({\mathbf x})~r^2\big(\cos{(2\theta)},~\sin{(2\theta)}\big)~{\mathrm d}^2
{\mathbf x}}
             {\iint I({\mathbf x}) ~W({\mathbf x})~r^2~{\mathrm d}^2{\mathbf x}} ~,
\end{equation}
\noindent where 
\begin{equation}
\label{eqn:ksbweight}
W({\mathbf x})=e^{-r^2/2r_g^2} ~.
\end{equation}
Although $\kappa$ is generally the desired quantity, and could be obtained in principle from
measurements of galaxy sizes \eqref{eqn:ksbd} or fluxes, this has proved difficult in practice,
because expectations for these quantities prior to lensing are unknown. On the other hand, while
galaxies have a natural dispersion of intrinsic ellipticities \eqref{eqn:ksbe}, they are (almost)
uncorrelated with each other in the absence of lensing, i.e.\ $\langle\varepsilon_i\rangle=0$. Any
correlation between the {\em observed} ellipticities of galaxies seen along adjacent lines of sight
arises because their light has traversed similar intervening large-scale structure
$\rho(\mathbf{x},z)$. In practice, corrections to measured ellipticities also need to be made for
the smearing of galaxies by the PSF, and for the differing susceptibilities of some galaxy
morphologies to an input shear. For more details of this procedure, see e.g.\ \citep{ksb}.

The observed shear can finally be transformed into convergence through their close relation in
Fourier space
\begin{equation}
\label{eqn:gammatokappa}
\tilde{\kappa}=
\frac{(\ell_1^2-\ell_2^2)\tilde{\gamma_1}+(2\ell_1 \ell_2)\tilde{\gamma_2}}
     {(l_1^2+l_2^2)}
\end{equation}
\noindent \citep{kaiser93}.  This is typically some amount, to reduce noise. 
Furthermore, like any scalar quantity extracted from a vector field, a
convergence signal can also be split into two independent components, 
$\kappa = \kappa^{E} + i~\kappa^{B}$ \citep{king01}.
The grad-like ``$E$-mode'' is the signal produced by weak lensing.
The curl-like ``$B$-mode'' is not produced by physical processes \citep[except at very low levels,
as described by][]{schneider02}, and therefore ought to be consistent with zero in the absence of
systematics. Usefully, it contains the same noise properties as the $E$-mode signal -- so it acts
as an independent realization of noise in the field, and any significant deviations from zero
alert to the presence of residual systematics (such as imperfect correction for the PSF).

\section{Observations and Data Reduction}
\label{sec:data}

\subsection{The COSMOS Data Sets}

Our data all cover the COSMOS survey field, a 1.64 deg$^2$ contiguous square, centered at
10:00:28.6, +02:12:21.0 (J2000) \citep{scoville07a}. The ground-based imaging was obtained in eleven 
mosaiced pointings
of the {\it Suprime-Cam} camera at the prime focus of the Subaru telescope on Mauna Kea
\citep{miyazaki02b}. These were taken on the 18th and 21st of February 2004, nights selected for
their excellent observing conditions: the mean seeing was $0.54\arcsec\pm0.03\arcsec$. The field
constitutes part of a larger weak lensing survey discussed, along with full details of the primary data
reduction pipeline, \com{in \citep[][Green et al.\ {\it in prep.}]{miyazaki07}. In fact, the relevant field 
in that survey covered a slightly larger area than the COSMOS field. 
The Subaru imaging was truncated when matching galaxy catalogs and was truncated {\it after} making 
convergence maps, to avoid edge effects associated with the
Fourier transform operations in equation~\eqref{eqn:gammatokappa}.}

Our comparison is made possible by the unique availability of deep, panoramic space-based imaging of
the COSMOS field \citep{scoville07b}. During HST cycles 13 and 14, 577 slightly overlapping
pointings were obtained from the {\it Advanced Camera for Surveys} ({\it ACS}) on board the Hubble
Space Telescope. Four dithered exposures at each pointing were stacked using the {\sc drizzle}
algorithm \citep{drizzle} to improve the native pixel scale of $0.05\arcsec$ and recover a final
pixel scale of $0.03\arcsec$. Full details of the primary data reduction pipeline for the HST images
are given in \citep{koekemoer07}.

It is important to emphasize that both the {\it ACS} and {\it Suprime-Cam} data exhibit
idiosyncrasies that present significant challenges for weak lensing analysis. For example, the
atmospheric seeing varied during the two nights over which the Subaru data were obtained; and the
distortions of the telescope's primary mirror under a gravity load were only passively corrected via
a look-up table as it followed the field. In a future ground-based experiment, such as LSST or
Pan-STARRS, seeing  variations could be normalized over a survey  by stacking a very large number of
short, independent  exposures taken over a long time period. Dome seeing could likewise be improved
with future  technologies. And while the telescope superstructure is particularly rigid at Subaru,
active correction  of the mirror support could undoubtedly improve future designs. Equivalently, the
sky background  seen from HST is affected by Earthshine that depends on the telescope pointing
\citep{leauthaud07}. The HST PSF also varies over time due to thermal fluctuations during each
low-Earth orbit  \citep{jee07, rhodes07}. Finally, the charge transfer efficiency of the {\it ACS}
CCD detectors had been  significantly degraded by high energy particles by the time the COSMOS data
were obtained, and worsened during the observing window \citep{rhodes07}. None of these problems
are inherent to all space-based observations: future missions might minimize or eliminate all three
effects by adopting a regular observing pattern, orbiting the Lagrange point L2, and using
radiation-hardened CCDs. However, the weak lensing analysis of existing space-based data is indeed
compromised by the extent to which such hardware variations can be modeled. In this sense, our
comparison is actually more informative than one based on simulated data that reproduces only
idealized and mean instrumental characteristics. 

The relevant characteristics of the two data sets are summarized in Table~\ref{tab:observations},
including limiting depths for a point source at $5\sigma$, in a $3\arcsec$ aperture from the ground 
and a $0.15\arcsec$ aperture from space \citep{capak07}. 
In addition to these images, the COSMOS field has been observed across all wavelengths from radio to
$X$-rays. Of particular relevance here are ($i$) deep $X$-ray observations by XMM
\citep{hasinger07}, which can be used to locate massive structures via thermal emission from hot
gas; and ($ii$) multicolor optical and near-IR imaging campaigns from the Subaru, Canada France
Hawaii, Cerro Tololo and Kitt Peak telescopes, which provide 15 additional bands and photometric
redshifts \citep{capak07, mobasher07}. The photometric redshift estimation code uses Bayesian priors
based on an adopted luminosity function, and includes reddening based on both Galactic and Calzetti
extinction laws. The results were calibrated using 868 galaxies in the field brighter than
$i^\prime=24$ and with spectroscopic redshifts. For galaxies closer than $z=1.2$, the rms scatter in
$(z_{phot} - z_{spec})\, /\, (1 + z_{spec})$ is 0.031.

\begin{deluxetable}{lcc}
\tabletypesize{\footnotesize}
\footnotesize
\setlength{\tabcolsep}{0.1in}
\tablecaption{Survey Characteristics}
\tablewidth{0pc}
\tablehead{& \colhead{Ground} & \colhead {Space}}
\startdata
Instrument                  & Subaru/{\it Suprime-Cam} & HST/{\it ACS}   \\
Primary aperture            & 8.2m                     & 2.4m            \\
Exposure time               & 5 $\times$ 360s          & 4 $\times$ 507s \\ 
Total survey time           & 5 hours                  & 325 hours       \\ 
Filter                      & $i^\prime$               & F814W           \\
Limiting AB magnitude       & 26.2                     & 26.6            \\
Field of View               & 2.14 deg$^2$             & 1.67 deg$^2$    \\
Pixel Scale                 & 0.202$\arcsec$           & 0.03$\arcsec$   \\
Point Spread Function       & 0.68$\arcsec$            & 0.12$\arcsec$   \\
\enddata
\label{tab:observations}
\end{deluxetable}

\subsection{Object Detection}

Objects were detected in the Subaru images using {\sc hfindpeaks} from the {\sc imcat}
package\footnote{Nick Kasiser's {\tt imcat} software package is available from  {\tt
http://www.ifa.hawaii.edu/$\sim$kaiser/imcat/}}. This finds the centroid and scale size $r_g$ that
maximizes the peak S/N of the image after smoothing with a Gaussian. The code also returns the
half-light radius, $r_h$, of each galaxy. Galaxies were initially detected to magnitudes fainter
than those for which it is possible to accurately measure shapes. To reduce noise in the final
analysis, weights were given to each galaxy, and galaxies with a detection S/N$<14$ were removed
from the catalog altogether. The resulting surface density is $n_\mathrm{gal}=42$ galaxies arcmin$^{-2}$.


Objects were detected in the {\it ACS} images using SExtractor \citep{BertinArnouts} in a dual
Hot/Cold configuration \citep{leauthaud07}, designed to identify both large and small objects while
avoiding fragmentation of the former, or merging  of the latter. The SExtractor centroids were then
improved, and the best-fitting scale size was selected, via an iterative process during shape
measurement. Galaxies smaller than $d=0.11\arcsec$ or fainter than $S/N\approx20$ were removed 
from
the catalog, and a weighting scheme was applied to faint galaxies as a function of their detection
S/N \citep{leauthaud07}. Note that an absolute calibration of the S/N was difficult to determine in
practice, because flux in adjacent pixels becomes correlated during DRIZZLE. The S/N cut corresponds
approximately to a limiting magnitude $F814W(AB)<26.5$ for a point source. The resulting surface
density is $n_\mathrm{gal}=71$ galaxies arcmin$^{-2}$, with a median redshift $z_\mathrm{med}=1.2$.

\subsection{Shear Measurement}
\label{sec:Shear}

Because the image characteristics of the two data sets are quite different, we adopted separate
methods to measure galaxy shapes, remove PSF effects, and ultimately obtain the weak lensing shear
signal. Each of these methods has been optimized for the respective data sets, so our comparison
will necessarily incorporate the limitations of each pipeline. We believe this is in the spirit of a
fair comparison of ground versus space. To minimize any differences arising entirely from the
algorithms themselves however, we have intentionally adopted related methods from the same
generation of software development and codes that have been well-tested.  Although newer shear
measurement methods \citep{k2k, im2shape, dahle02, shapelets2, bj02, hirata03, kuijken06,
shapelets4, reiko} may offer improved performance, none has yet been sufficiently tested  across
both observing regimes.

The Subaru images were analyzed with the \citep{hamana03} implementation of the widely used
\citep[][hereafter KSB]{ksb} shear measurement method. This particular implementation is a
derivative of the ``LV'' pipeline tested in the Shear TEsting Program
\citep{step1,step2}.

The HST images were analyzed with the \citep[][hereafter RRG]{rrg} shear measurement method. This is
a perturbation of the KSB method for space-based data. It calculates the same quadrupole moments,
but corrects them individually for the effects of convolution with the PSF, and only in the final
stage takes the ratio~\eqref{eqn:ksbe}. This is necessary because the small and cuspy
diffraction-limited PSFs otherwise introduce divisions by very small (and noisy) numbers. RRG been
applied to HST {\it WFPC2} \citep{rrg}, {\it STIS} \citep{rhodes04}, and {\it ACS} data
\citep{massey07a}. The {\it ACS} pipeline was thoroughly tested on simulated images during the
creation of the COSMOS catalog \citep{leauthaud07}, and also for a continuation of STEP using
simulated space-based images. 


\section{Statistical Applications}
\label{sec:analyses}

\subsection{Shear-shear comparisons}\label{sec:shearbyshear}

We shall now compare the global properties of our ground- and space-based shear catalogs, to  
determine the depth (and galaxy surface density) at which reliable shear measurement is possible
from the ground. This will be relevant for many statistical applications, including measurements of
the angular shear-shear correlation function that are typically used to constrain cosmological
parameters. In such analyses, where statistical noise is reduced by averaging over many lines of
sight, the key issue is the reliability and level of residual systematics in the shear measurement.

We asses the performance of the ground-based shear measurements against those from {\it the same
galaxies} in space-based data, making the necessary but reasonable assumption that the shapes are
much more reliable when measured from the much higher resolution images with a smaller PSF. Such a
comparison is clearly only possible for the subset of galaxies contained in both catalogs. The two
quantities of interest will be linearity in the comparison (the slope of the shear-shear
comparison is equivalent to the STEP ``calibration bias'' parameter $m$) and the scatter (which
represents the combined shear measurement noise from both HST and Subaru, plus any systematic
effects).

We match galaxies whose positions agree to within 1$\arcsec$, and produce a common catalog 
containing $n_{gal}=32$ galaxies arcmin$^{-2}$. Many objects in the Subaru galaxy catalog without
matched counterparts in HST galaxy catalog have half-light radii on the limits of seeing and are likely to have been
revealed as stars by the higher resolution data; in any case, the omitted galaxies had below-average weights in
the Subaru catalog. The remaining unmatched objects are a combination of noisy/skewed galaxies with
offset centroids, or galaxies that lie in regions of the HST images masked because of scattered
light from nearby bright stars. For the following tests tests, we shall ignore the weights on remaining
galaxies, and treat all objects equally.

\begin{figure}
\centerline{\epsfig{file=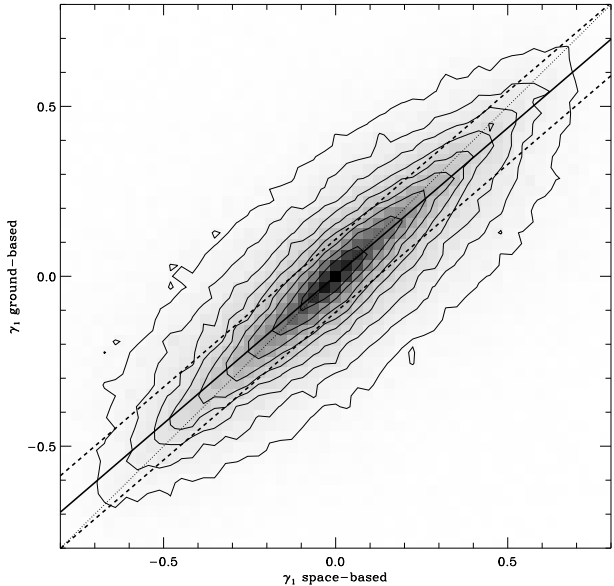,width=3.2in}}
\caption[]{Comparison of shear measured from galaxies seen in both Subaru and Hubble Space Telescope images.
The greyscale shows the number of galaxies with different shear measurements. The outer contour
includes 90\% of the galaxies, and successive inner ones include 10\% fewer. 
The solid line is the least squares linear relation.
Its slope of 0.87 indicates that shears have been underestimated in the ground-based analysis, 
or that the catalog is still partially contaminated by stellar sources.
This value is insensitive within 0.01 to the reintroduction of galaxy weights.
Furthermore, the non-Gaussian
wings of the scatter extend well beyond the rms error of 0.16, shown as dashed lines.}
\label{fig1}
\end{figure}

Figure~\ref{fig1} shows a comparison of the shear signal for the matched galaxies. Since errors  are
present in both axes, to calculate the best-fit linear relationship we adopt a least squares  
method that minimizes the perpendicular distance to the best fit line (instead of one that assumes 
one variable is error-free). Ideal shear measurement from both instruments would yield  a best-fit
slope of unity. \com{There will inevitably be a small amount of scatter, because the weight function~\eqref{eqn:ksbweight}
is not necessarily the same size $r_g$ in the ground- and space-based analyses.
In practice, we find a best-fit slope of 0.87, indicating that the shears have been
underestimated from the ground. The measurement noise is also
problematic, with $\sigma_\gamma=0.16$ per component (perpendicular to the best-fit line; 
note that this does not include intrinsic source ellipticity variance because the galaxies are matched) and
a skewed, non-Gaussian distribution of outlying shear estimates} that would render a cosmic shear analysis less stable. 

\begin{figure}
\centerline{\psfig{file=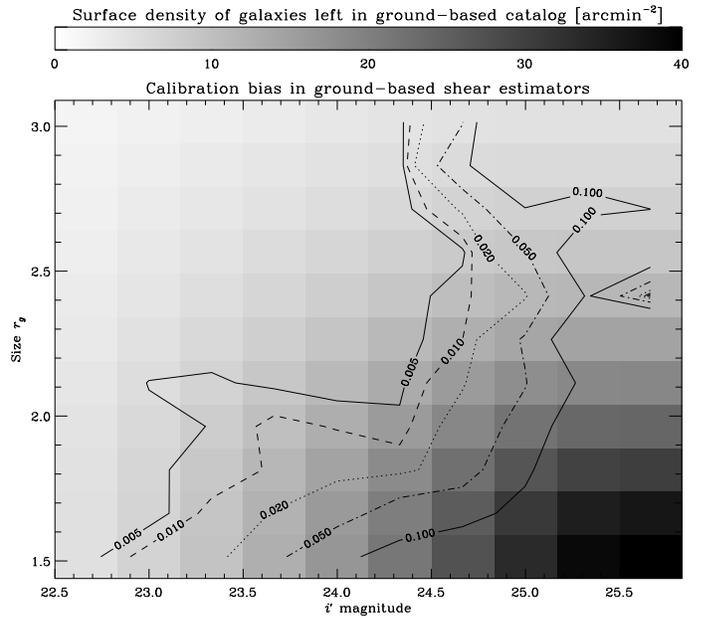,width=3.2in,angle=90}}
\caption[]{Relative calibration between shear measurements from galaxies in Subaru and Hubble Space Telescope data, 
for galaxies of different sizes and magnitudes. 
The contours show deviations from a slope of unity in figures~\ref{fig1} and
\ref{finalcuts}, which would have been ideal. For faint galaxies, these deviations tended to be an
underestimation in the Subaru pipeline relative to HST. There is some evidence that shears are
overestimated in large, bright
galaxies, although the small number of these objects means that the extrapolation is less certain.
The calibration biases are calculated locally, for
galaxies only in a given cell of \{size,magnitude\} space.
On the other hand, the grey-scale shows the cumulative number density of galaxies $n_{gal}$  that 
would remain in a ground-based catalog, were cuts to be applied at the local values 
(i.e.\ including all larger and brighter galaxies).}
\label{fig:calibration}
\end{figure}

\begin{figure}
\centerline{\psfig{file=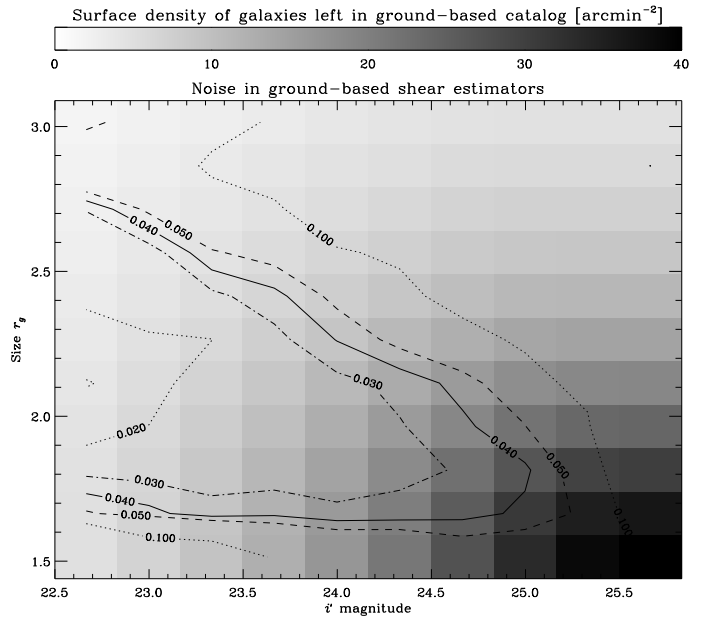,width=3.2in,angle=90}}
\caption[]{Combined noise from shear measurements of galaxies matched in catalogs from Subaru
and Hubble Space Telescope data. The contours show $\sigma_\gamma$ as a function of galaxy size and
magnitude. As in figure~\ref{fig:calibration}, these are calculated only for galaxies with that
particular size and magnitude. The contours close at the top merely because there are very few
large, faint galaxies, so the rms scatter increases. 
The grey scale again shows the total number density of available
galaxies.}
\label{fig:noise}
\end{figure}

The overall performance in figure~\ref{fig1} is a superposition of good shears from bright and (in
particular) large galaxies, plus smaller objects that cause most of the bias and scatter. Indeed,
systematic errors could be completely eliminated by using only the very largest galaxies. However,
the statistical noise in a cosmic shear analysis of shear-shear correlation functions scales as
$\sigma_\gamma/\sqrt{n_\mathrm{gal}}$. An optimal strategy for any particular ground-based survey
will involve catalog cuts requiring a trade-off between systematic and statistical errors. However,
the optimal cuts will vary as a function of survey area and depth. To produce a result of general
interest, we therefore show in Figures~\ref{fig:calibration} and \ref{fig:noise}, the resulting
calibration bias, scatter and galaxy density for a range of possible cuts in galaxy size and
magnitude. 

\begin{figure}
\centerline{\epsfig{file=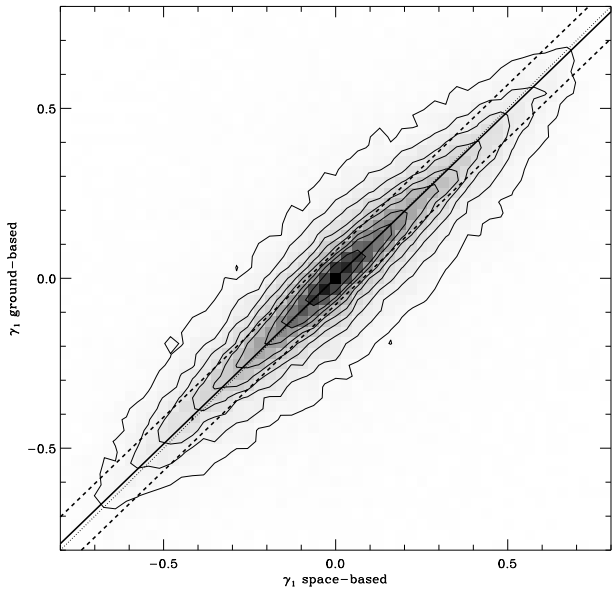,width=3.2in}}
\caption[]{As for figure \ref{fig1}, but for the subset of galaxies brighter than $i^\prime=24.5$
and larger than $r_h=1.8\arcsec$.
The total least squares slope is 0.97, implying an almost unbiased recovery of the shear signal
from Subaru, and the data is better and more symmetrically enclosed within the rms scatter of 0.11.}
\label{finalcuts}
\end{figure}

A simple result emerges from Figure~\ref{fig:calibration}. It is noticeable from the horizontal and
vertical contours that size and magnitude cuts seem to neatly parametrize independent sources of
error. Using existing shape measurement methodology, shear can be measured from galaxies brighter
than $i^\prime=24.5$ and larger than $r_h=1.8$, with measurement noise $\sigma_{\gamma}\simeq
0.03$ and a calibration bias less than 3\% \com{(and only 1\% with galaxy weighting)},
which is acceptable for competitive constraints from future surveys \citep{refregier04}. This leaves
a surface density of $n_\mathrm{gal}=15$ galaxies arcmin$^2$ from the ground, with a median redshift
of $z_{\rm med}=$0.8. The comparison with space-based data for these cuts is shown in
figure~\ref{finalcuts}. 

Note that we have not been able to test the reliability of space-based shear measurements using this
method, nor even considered the population of small galaxies resolved only from space. Without data
even better than {\it ACS} imaging to compare to, we resort to simulations. The full RRG pipeline
was calibrated against simulated images by \citep{leauthaud07}. However, of the 71 galaxies
arcmin$^{-2}$ successfully in real {\it ACS} images, only the brightest 40 could be used by 
\citep{massey07a} to minimize the $B$-mode signal and overcome problems of CCD charge transfer
inefficiency (CTI). This limitation clearly needs to be overcome: perhaps via a CTI correction
algorithm like that developed for {\it STIS} by \citep{cte}, and radiation-hardened detectors in
future telescopes.

\subsection{Survey speed}\label{sec:surveyspeed}

Our ground versus space shear comparisons have important implications when it comes to
considering the optimal  approach for measuring shear for cosmological applications.
Although the {\it \'etendue} of instruments can be expected to increase both on the ground and in space,
we will base our discussion on the imaging depth and fields of view of our HST and Subaru surveys.

An important criterion is what can be accomplished in a given amount of observing time; the HST and
Subaru requirements for our comparison are summarized in Table~\ref{tab:observations}).  HST 
overheads
approximately tripled the on-source exposure time, and, at Subaru,  high quality imaging was
secured during what might be considered a fortuitous observing window. Noting that \citep{bacon01}
found images with seeing worse than $\sim 0.8\arcsec$ of little use for weak lensing analysis, 
coupled with observational visibility, it
seems reasonable to incorporate a factor of at least four inefficiency for a generic survey: even for a
superb ground-based facility such as Subaru, on an excellent site such as Mauna Kea\footnote{{\tt
http://www.cfht.hawaii.edu/Instruments/Imaging/Megacam/ observingstats.html}. Future surveys such as
Pan-STARRS and LSST, which plan to co-add  many short exposures with independent PSFs, may 
achieve
near-uniform image quality  by rejectng a certain fraction of exposures. But the relevant figure of
merit is still the fraction of time spent with seeing better than  $0.8\arcsec$.}.  Based on its
superior field of view, Subaru is then $\sim$ 24 times  faster than HST in  useful mapping  speed. As,
to  first order, the signal to noise in statistical analyses increases as $\sqrt{n_\mathrm{gal}}$,  for
a fixed survey lifetime, this  corresponds to a $\sim5$-fold  improvement in signal to noise. 

This simplistic analysis is of course mitigated by the higher resolution available from space. We next
insert the gain in surface density, viz  71 galaxies arcmin$^{-2}$ resolved in our space-based imaging 
c.f.\ 15 arcmin$^{-2}$ from the ground. We will assume that the additional, small galaxies have a similar
distribution of intrinsic 
ellipticities as the larger ones \citep[c.f.][]{snap2, leauthaud07} and that the measurement noise on an 
average survey galaxy is constant (since the size distribution of resolved galaxies compared to the PSF 
size is roughly independent of the PSF size). Incorporating this increased background surface density,
the ground-based gain per unit time drops to only a factor of 2.3. 

\begin{figure}
\centerline{\psfig{file=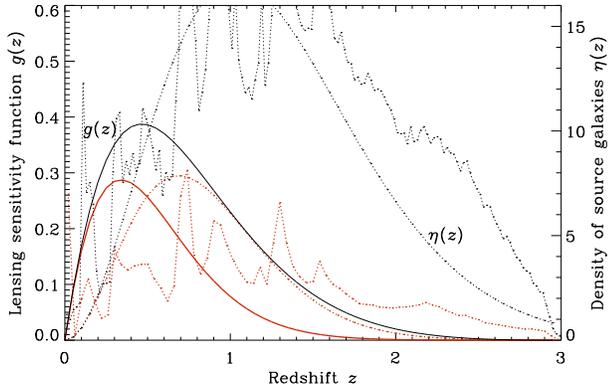,width=2.in,angle=90}}
\caption[]{The dotted lines show the redshift distribution of source galaxies from 
Hubble Space Telescope (black) and Subaru (red) imaging surveys, for the catalog cuts used in figure~\ref{finalcuts}.
The jagged lines show the measured photo-$z$s, and the smooth curves assume a simple parametric 
form for the background galaxy redshift distribution from \citep{smailzdist}, 
with $\alpha=2$, $\beta=1.5$, $z_{\rm med}=$0.8 or 1.2, and an overall
normalization to reproduce the observed number density of galaxies.
The solid lines show the corresponding lensing sensitivity functions calculated from the analytic curves.
Thees lie always in front of the
source galaxies but are notably higher for a space-based survey, particularly at redshifts greater
than 0.5. Extending this redshift coverage is crucial for 3D tomographic analysis techniques
to measure the growth of structure.}
\label{fig:zsensitivity}
\end{figure}

Equally important to the increased density of galaxies are their higher redshifts. Distant galaxies are 
more
sensitive to
low-redshift lenses, and sensitive to more total lenses. The shear signal grows proportionally to the
median source redshift $z_{\rm med}^{0.6-0.8}$ \citep{jainseljak97}. With the redshift distributions
for galaxies shown in figure~\ref{fig:zsensitivity}, the total gain in signal to noise for a 2D weak
lensing survey conducted from Subaru over one conducted with HST is only about 1.7.

Perhaps the most important advantage of space is the increased redshift range of resolved galaxies. 
This better enables their stratification into redshift bins for tomographic (3D) analyses. Deep infrared 
imaging, needed for accurate photometric redshifts, is also likely to remain the province of space-based
observatories.  Tomographic techniques can tighten the constraints on cosmological parameters $
\Omega_{\mathrm
M}$ 
and $\sigma_8$ by a factor of at least three \citep{massey07a} and potentially as much as 
five \citep{heavens06}. 
Further advantages of these techniques includes the elimination of unwanted signal from 
adjacent galaxies' intrinsically-correlated shapes \citep{king03, heymans03}. While wide-field ground-based
instruments may therefore yield significant improvements for  Dark Energy Task Force ``Stage 3'' 
surveys, 
advanced analysis techniques for ``Stage 4'' surveys will realistically be possible only with space-based
facilities.
These will bring new scientific opportunities, cross-checks for systematics, and greater efficiency.

\section{Mass Maps and Halo Detection}
\label{sec:Maps}

\begin{figure*}
\centerline{\epsfig{file=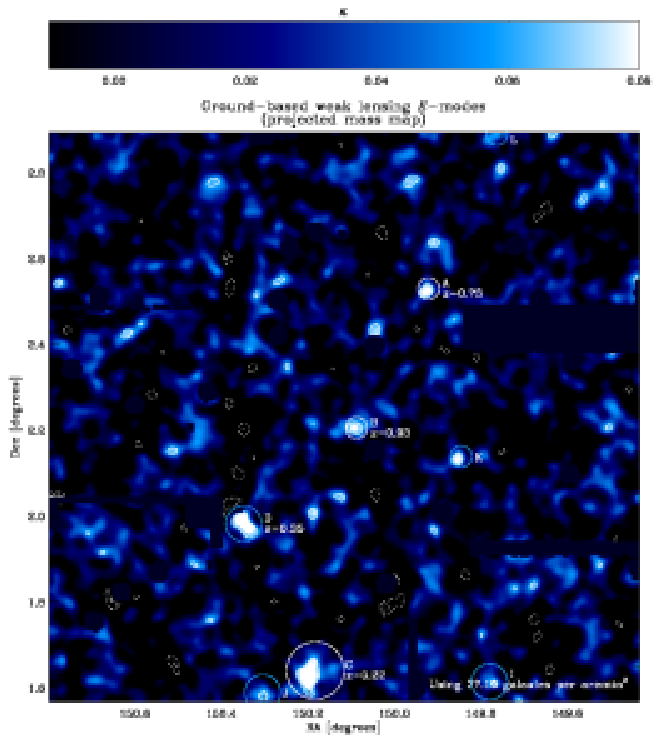,width=3.5in,angle=0}\epsfig{file=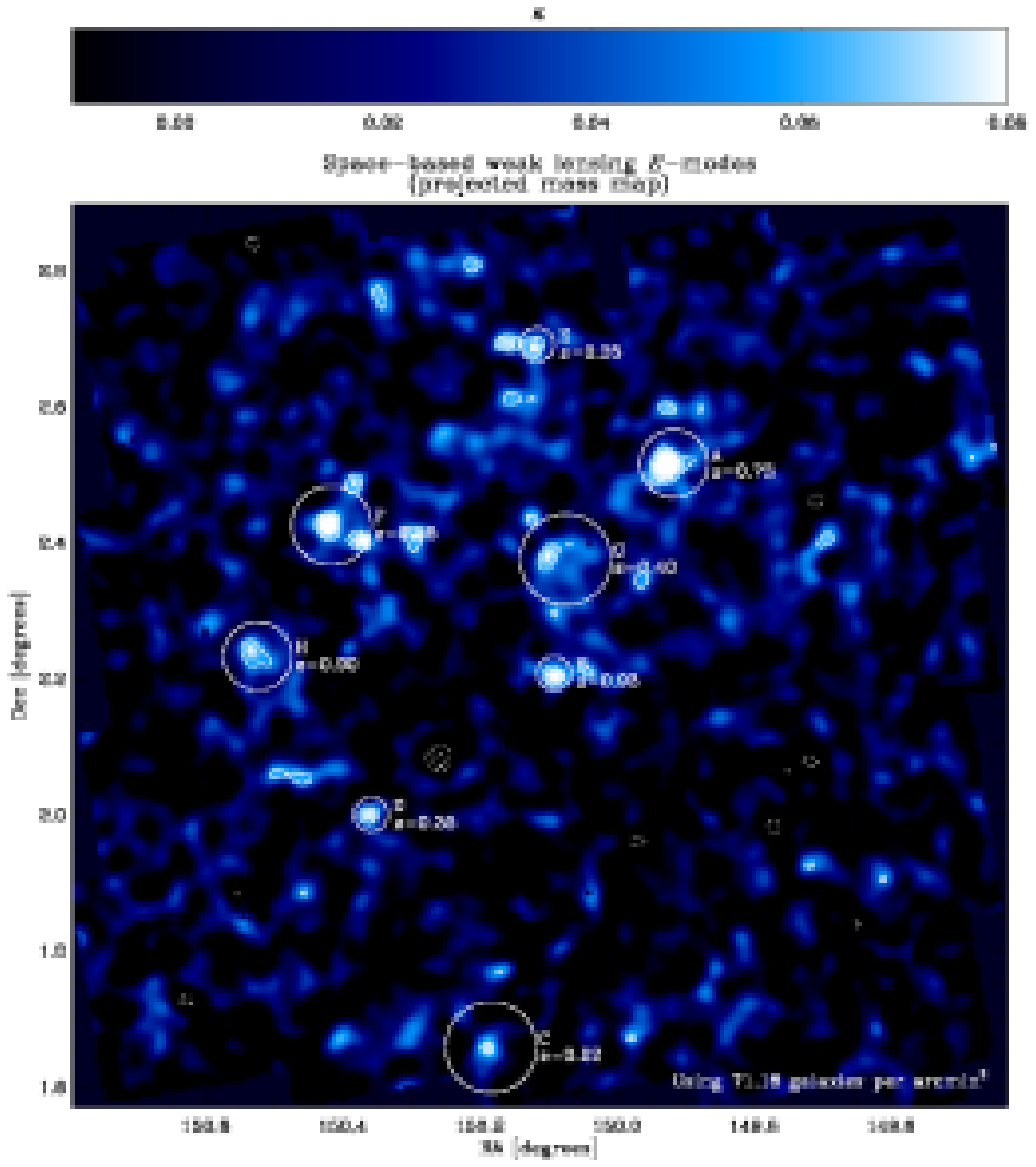,width=3.5in,angle=0}}
\caption[]{Convergence $E$-mode maps from the Subaru (left) and
Hubble Space Telescope (right), \com{after smoothing by a $1\arcmin$ Gaussian kernel.
The data presented in the left panel are identical to that in figure 13 of \citep{miyazaki07}, except that the
field has been slightly truncated to match the right panel.
Convergence is proportional to the total projected mass along a line of sight, 
modulated by the
lensing sensitivity function \eqref{eqn:sensitivitydefn} plotted in figure~\ref{fig:zsensitivity}. 
Contours are drawn at detection significances of $3\sigma$, $4\sigma$ and $5\sigma$, with dashed lines for underdensities.
Clusters A, B, C and D are detected in both maps.
Other peaks E-L are only detected in one of the two.
White enclosing circles denote clusters deemed ``secure'' by the rigorous standards of \citep{miyazaki07}, and cyan circles
denote ``unsecure'' clusters. The size of the circles shows the size of the smoothing kernel that maximises detection significance,
enlarged by a factor of 2 for clarity. 
}}
\label{emode}
\end{figure*}

\begin{figure*}
\centerline{\epsfig{file=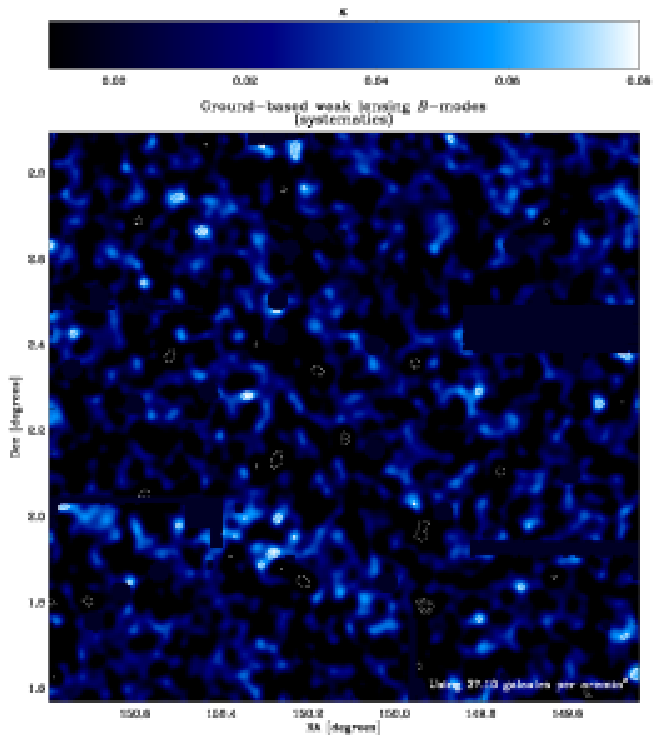,width=3.5in,angle=0}\epsfig{file=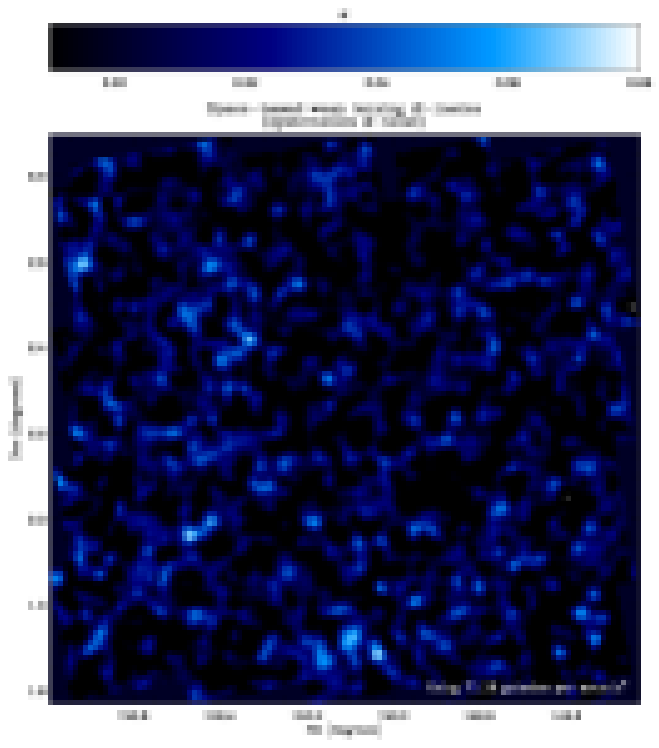,width=3.5in,angle=0}}
\caption[]{Convergence $B$-mode maps from Subaru (left) and Hubble Space 
Telescope (right) data. This is not
produced by physical gravitational lensing, so deviations from zero include a combination
of spurious effects from e.g.\ imperfect PSF correction, plus a realization of statistical noise.
\com{The smoothing scales, color ramps and contours are identical to those in figure~\ref{emode}.}}
\label{bmode}
\end{figure*}

We now investigate the reconstruction of maps of the mass distribution (figures~\ref{emode}  and
\ref{bmode}), and the detection of individual mass peaks. The mass and redshift distribution
$N(M,z)$ of several thousand lensing-selected clusters could be used to constrain cosmological
models \citep[][Green et al.\ {\it in prep.}]{hamana03, wang04}. Additionally, the physical properties of the 
dark matter
particles can be investigated by comparing the detailed distribution of dark matter with that of
baryons \citep{clowe06, dmring}. The key issues will be the angular resolution of reconstructed mass
maps, as well as the mass and redshift range in which halos can be successfully detected. We treat
this as an independent experiment from the previous section, beginning the comparison of ground- and
space-based data afresh. In particular, we do {\it not} cut the Subaru data  to the shallower depth
discussed in \S\ref{sec:shearbyshear}, to eliminate the last few systematic biases. 
The intent is  not to align our two comparisons but rather to optimize each analysis as an 
independent experiment -- as would be the case if either were being undertaken as a self-contained 
survey.

Unfortunately, even with the unprecedented investment of HST time for the 
COSMOS survey, we can expect the number of lensing-detected structures in this finite field 
to be modest. At the Subaru depth, a surface
density of $\sim$5 halos deg$^{-2}$  \citep{miyazaki07} implies only 
around eight halos are likely to be found in the COSMOS field. 
Thus we recognize in this comparison that the statistical significance 
of our results will be quite limited.

\subsection{Residual Systematics}

First, we consider the $B$-mode signal. As discussed in \S\ref{sec:theory}, the $B$-modes act as an
independent realization of noise in the mass map, and locally highlight any problems with the
correction for PSF or other effects peculiar to the (two very different) instruments.
Unsurprisingly, a visual inspection of Figure~\ref{bmode} shows that the $B$-mode signal is
significantly lower in our space-based data, with fewer $B$-mode peaks. The overall noise level is
reduced, and holes arising from masked foreground stars are also smaller and less frequent. In the
ground-based maps, these create additional edges that lead to spurious effects during the Fourier
transforms required by equation~\eqref{eqn:gammatokappa}. 
The extended gaps are caused by difficulties modeling the PSF near the edge of the field of view,
\com{and could be eliminated in future surveys by more conservative tiling strategies}.

%
%
\com{The southwest corner of the field has been
troublesome throughout our analysis. This pointing was observed in slightly worse seeing, so the
density of galaxies is reduced and the noise in the mass reconstruction is higher.}


\subsection{Halo Detection}
\label{sec:halos}

The higher surface density of background galaxies from space also improves the reconstruction of the
$E$-mode ``mass map''convergence field. The noise is lower and the angular resolution higher
(although to aid comparison, both panels in figure~\ref{emode} are smoothed to the same scale).
Several of the key features are qualitatively similar but we are struck by the significant differences
in the prominence of other mass peaks. 
\com{To evaluate the robustness of detections, we shall now employ an automated peak-finding
algorithm.}

\com{ Following \citep{miyazaki07}, we smooth the convergence maps by a Gaussian kernel of rms 
width 1$\arcmin$ and find local maxima with detection significance $\nu>4$ (assuming Gaussian errors 
on the shear measured within $0.7\arcmin$ cells on the sky equal to the dispersion of those galaxy shears). 
Five peaks (marked A, B, C, D and K in figure~\ref{emode}) are then identified in the ground-based data. 
However, two of these are near boundaries of the field mask. Imposing the rigorous restrictions discussed
by \citep{miyazaki07}, we find that only peaks A, B and C remain \citep[c.f.\ table~3 in ][]{miyazaki07}.  All 
three are also detected in a space-based lensing analysis, the 3D distribution of galaxies and as extended 
sources in  $X$-ray data \citep{hasinger07,finoguenov07}.} Assuming the mass-luminosity relation adopted 
by \citep{finoguenov07}, the detection threshold of this very deep $X$-ray data is well below that expected
for lensing up to redshift $\sim1$, so this acts as an ideal external arbiter (of course, $X$-ray 
mass-observable relations are somewhat uncertain). The properties of the three clusters are summarized 
in table~\ref{tab:mass} and demonstrate excellent agreement between the ground- and space-based
data using the formalism of \citep{miyazaki07}.

Cluster A (SJ J0959.6+0231) is the most massive structure inside the COSMOS field, easily detected
at many wavelengths. It appears to be in the process of a major merger, and has been studied
individually by \citep{guzzo07}, who also obtained a spectroscopic redshift of $z=0.73$.
Cluster B (SL J1001.4+0159) is associated with an $X$-ray peak
and overdensity of galaxies at $z_\mathrm{phot}=0.35$. There is a second set of galaxies at
$z_\mathrm{phot}=0.85$ within $2\arcmin$, which undoubtedly complicates the interpretation a
little, but our results are consistent with this high redshift projection being a minor
perturbation. Cluster C (NSC J100047+013912) is yet more local \citep[$z=0.22$,][]{miyazaki07} and 
appears large on the sky. Only part of this cluster is inside the region of HST imaging, so the space-based 
signal is significantly weakened, and the mass is potentially underestimated by HST.

\com{ To broaden our search, and test the limits of detectability, we additionally investigate the 
multi-scale procedure of \citep{hamana03}. For this, we smooth the convergence maps with Gaussian 
filters of rms width between 0.5$\arcmin$ and 4$\arcmin$, identifying local maxima inside the mask on 
each scale. For each peak with a detection signal to noise $\nu>4$ on any scale, we record $\nu$
and the smoothing scale that maximizes $\nu$. We also drop the restrictions on distance from the 
mask boundaries. This will increase the number of detected peaks, but at the expense of potentially 
introducing some spurious features. We then search for counterparts in the other data set, within 
3$\arcmin$ of detected peaks.}

With the above criteria, we identify four mass peaks common to both convergence maps (A, B, C and D).
Cluster D is within 3$\arcmin$ of an $X$-ray peak, and an overdensity of galaxies at photometric 
redshifts $z_\mathrm{phot}=0.93$.  This redshift is rather high for a lensing analysis, and it was flagged 
as ``unsecure'' by \citep{miyazaki07} because it is near a boundary in the image mask.
However, the tentative Subaru detection is strengthened by the confirmation from HST, and appears 
to be robust.

Peaks E, F, G and H (also marked on Figure~\ref{emode}) are seen only in the space-based map. The first three
correspond to extended $X$-ray emission from clusters with masses $M_{500}$ between $2-4\times10^{13}M_\sun$
\citep{finoguenov07}. Peak H is more massive ($M_{500}=1.8\times10^{14}M_\sun$), but is at very high
redshift. \com{All four of these peaks are real detections; however no counterparts within 3$\arcmin$
are seen in the ground-based map, even down to $\nu>$3. Most likely, this is because of their lower mass and
higher redshift \citep{hamana03}.} The detection of peak G
was prevented in the ground-based data by a bright foreground star.


\com{
Conversely, peaks I, J, K and L are detected only in the ground-based map.
Peaks I and J are real, but lie just outside the HST imaging.
There is an extended $X$-ray source at peak I, with unknown redshift, and 
a projection of two $M_{500}\approx2\times10^{13}M_\sun$ clusters at redshifts $z=0.40$ and $z=0.75$ at peak J \citep{finoguenov07}.
In both cases, there is a weak, $\nu<3$ signal in the HST data, from the wings of the cluster.
Peak K was detected in the \citep{miyazaki07} analysis, but again flagged as
``unsecure'' because it is near the edge of a pointing.
It does align with a slight, $\nu<3$ detection in the space-based map, 
but there is no $X$-ray counterpart. This may be a spurious peak with chance coincidence, or perhaps a very
distant object.
Peak L appears to be spurious: such noise artifacts are more common near the edge of the field. 
}

In summary, to the extent that we can draw conclusions from such a small sample, there
is very good agreement between the primary halo catalog drawn from the ground-based
data and that independently found from the space-based data. Additional halos of
lower mass and higher redshift are seen in the space-based catalog, and those located uniquely in the
ground-based data can be understood in the context of either being outside the space-based
region or close to its periphery.

\begin{deluxetable*}{cccccccccccc}
\tabletypesize{\footnotesize}
\footnotesize
\setlength{\tabcolsep}{0.07in}
\tablecaption{Cluster Masses}
\tablewidth{0pc}
\tablehead{
\colhead{Cluster}   &\colhead{RA}  &\colhead{Dec} &\colhead{Redshift}  &  \colhead{XMM mass} & \colhead{HST mass}  & \colhead{Subaru mass} \\ 
   & &   & & \colhead{10$^{14} M_{\sun}$} & \colhead{10$^{14} M_{\sun}$} & \colhead{10$^{14} M_{\sun}$}}
\startdata
 A  & 149.917 & 2.515   &  0.73 & $1.90\pm0.05$ & 23$^{+13}_{-8}$  & $13^{+33}_{-9}$   \\
 B  & 150.359 & 1.999   &  0.35 & $0.10\pm0.01$ & 9$^{+7}_{-4}$    & $17^{+13}_{-7}$   \\
 C  & 150.184 & 1.657   &  0.22 & $1.01\pm0.02$ & 17$^{+17}_{-9}$  & $55^{+55}_{-27}$ 
\enddata
\label{tab:mass}
\end{deluxetable*}


\subsection{Halo Mass Estimation}

\begin{figure}
\centerline{\epsfig{file=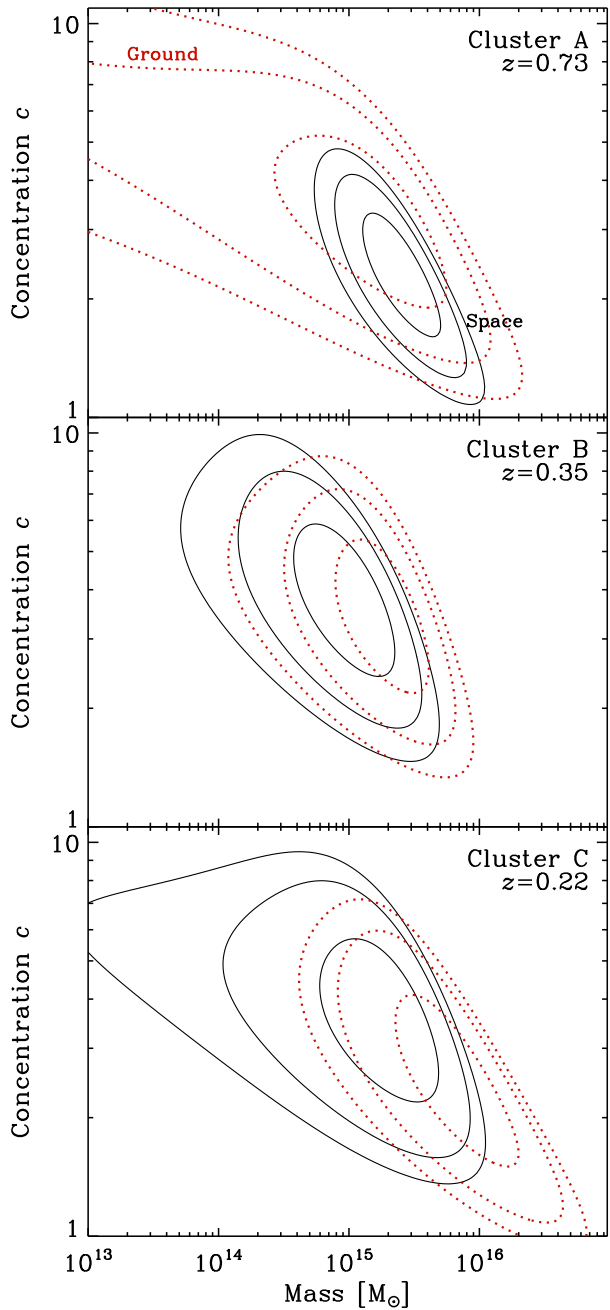, width=80mm}}
\caption[]{Best-fit mass and concentration index of three clusters in the COSMOS field, assuming
NFW radial mass profiles. The contours show the 68\%, 95\% and 99\% confidence regions obtained
from Subaru data (dotted red) and Hubble Space Telescope data (solid black).}
\label{masses}
\end{figure}

We now attempt to measure the total mass of each of the three halos (A-C) securely detected from both
the ground and space.
We assume that the clusters have an NFW density profile
\begin{equation}
\rho(r) = \delta_c ~\rho_c/(r/r_s)(1+r/r_s)^2 ~
\end{equation} 
\noindent \citep{nfw}, where $\delta_c$ is a function of the cluster's concentration $c$
and scale size $r_s \equiv r_{200}/c$, \com{and $r_{200}$ is the radius within which the mean density
is 200 times the critical density.} 
The shear profile of an NFW cluster is derived by \citep{king01}. 
We perform a maximum likelihood fit to the log(mass) and concentration parameters, using the shear 
measurements from all galaxies within 10$\arcmin$ of the \com{peak convergence signal}, averaged in radial bins of 0.5$\arcmin$.
\com{
It has been variously noted (J.\ Berg\'e, S.\ Paulin-Henrikkson, private comm.) that fitting noisy data 
of individual clusters with an NFW profile does permit large (and therefore massive) models with 
unnaturally low concentration values.
To counter this effect, we impose a concentration prior, using the lognormal distribution found 
for all haloes in the {\it Millenium Simulation} as a function of mass
by \citep[][equations~(5), (6) and figure~6]{neto}.
The resulting likelihood surfaces are shown in figure~\ref{masses}, 
with the effect of the prior being to close the bottom of the contours.
Table~\ref{tab:mass} lists the best-fit masses and 68\% confidence limits after marginalizing over
concentration between $1<c<10$.
}


Although our common sample is small, there is an encouraging agreement between the detailed
properties of the clusters recovered from the ground and from space. For the higher redshift cluster A,
our space-based data does put significantly tighter constraints on the mass and
concentration than our ground-based data. However, for the lower redshift clusters B
and C, the results are satisfyingly similar. We note again that cluster C is partially outside the
HST imaging. Since shears are only measured around one half of the cluster, 
the \com{statistical errors} in the space-based analysis are \com{larger and its mass could be underestimated}.
Certainly, for massive clusters with redshifts 
$0.2\simlt z\simlt 0.5$, it appears that our ground-based depth and resolution is adequate. The main benefit of 
space-based imaging is in the measurement of lower mass halos and higher redshift clusters, plus the 
increased resolution to further investigate the distribution of their masses.

\section{Discussion}\label{sec:conc}

We have performed parallel weak lensing analyses of Subaru and Hubble Space Telescope imaging in 
the
COSMOS field. Our comparisons of the observed shear and convergence signals have revealed a 
number of
issues, and suggest that such a study with real data usefully complements the independent
approach based on blind analyses of simulated data \citep{step1, step2}. 

For statistical ``cosmic shear'' analyses, shear measurement with an existing ground based
telescope, using existing measurement techniques, can be achieved with less than 1\% bias relative
to higher resolution space-based data, for a galaxy surface density of 15 arcmin$^{-2}$. 
One limitation of our approach is that we cannot check the performance of our space-based analysis
on the additional, small galaxies.
At first sight the factor of $\sim$3 shortfall in surface density seems inconsequential given the
lower cost and improved areal mapping speed of existing ground-based cameras
such as SuPrime-Cam. However, accompanying the brighter Subaru limit is a
reduction in survey depth and hence the redshift distribution of background sources.
More distant sources contain a larger signal, and a narrower range of redshifts also hinders
tomographical tests \citep{bacon05,massey07b}, which tighten cosmological parameter constraints
significantly. 

A key issue is whether this limiting depth is a fundamental one for all future
ground-based cameras. PanSTARRS, VST, and even LSST each have significantly
smaller primary mirrors than Subaru, so achieving even the $S/N$ discussed here
would require formidable exposure times. Most importantly, the deep infrared imaging
that is required for photometric redshifts to enable tomographic analyses is likely to be
difficult over large survey fields from the ground, because of increased sky background.
Recent weak lensing analyses are limited at roughly the same level by uncertainty in
galaxy shape measurement and photometric redshift estimation. 

Statistical measurements from the ground are also hindered by variable atmospheric seeing. Past
experience has taught the authors that data collected in seeing worse than $0.8\arcsec$ is
of little use for weak lensing analysis. The apparently rapid speed of data collection for our Subaru
data belies the time spent waiting for better seeing, even with the excellent atmospheric
conditions above Mauna Kea and the well-controlled dome seeing of Subaru. For this small-scale
survey, we obtained exceptional quality imaging during a fortuitous observing window. The relevant
quantity for larger-scale surveys in the future will be the time-averaged seeing quality, and the
fraction of time spent with seeing better than $0.8\arcsec$. This is particularly true for surveys
like Pan-STARRS and LSST, that plan to adopt a strategy of co-adding many shorter exposures. 
Their advantage is that the stacked images will achieve a near-uniform image quality, by virtue of the
independent PSFs in each short exposure. This can then be tuned to the required image quality by
rejecting a certain fraction of exposures.

Variable seeing conditions is also of concern for the reconstruction of mass maps (c.f.\ Green et al.\ {\it in 
prep.}). Difficulties in
the analysis of one pointing in the SW corner of the Subaru map result in a patchy recovery of
large-scale structure; with \com{more noise and a lower range of probed redshift}
in certain regions. 


Most importantly, \com{however, the four most massive clusters out of eight detected from space are also detected from the ground --
with one intriguing additional signal and two more confirmed clusters just ouside the field of view observed from space.
Reassuringly, the three clusters conservatively deemed ``secure'' by the independent analysis of \citep{miyazaki07} 
have now been confirmed via space-based weak lensing and $X$-ray observations.
The physical properties of the four massive halos in common (A, B and C) are remarkably consistent whether derived from 
from ground- or space-based weak lensing. The measured masses and radial profiles of these clusters are consistent and, for
the lower redshift clusters, the error bars are comparable.}
A Dark Energy Task Force ``Stage 3'' survey from the ground appears eminently feasible. 

 
A wide-field space-based platform would open up many new applications. Very important for statistical
applications is the increased redshift range of resolved background galaxies. These not only contain
a larger shear signal, but also more readily split into redshift bins for tomographic analysis.
Three-dimensional analysis techniques will tighten constraints on cosmological parameters by factors
of $3-5$, and directly measure quantities that depend upon the properties of dark energy, like the
growth of structure over cosmic time and the redshift-distance relation. They will also eliminate
sources of error due to the intrinsic correlations of galaxy shapes. Sufficiently good photometric
redshifts require deep, wide-field near-IR imaging, and these are also realistically possible over
large surveys only from space. Recent weak lensing analyses with relatively shallow near-IR coverage
like \citep{massey07a} are limited to roughly the same degree by uncertainty in galaxy shape
measurement and photometric redshift estimation. Full implementations of cross-correlation
cosmography will almost certainly require deep near-IR imaging from space. Such advanced techniques
will become particularly important as ground-based surveys expand to encompass the entire observable
sky.

The increased surface density of galaxies resolved from space also improves maps of the mass
distribution. \com{As shown in figures~\ref{emode} and \ref{bmode}, the statistical noise and systematic 
contamination in the $B$-mode are significantly reduced.
Eight clusters are detected in the COSMOS $E$-mode signal without any contamination from spurious peaks.
With the increased mass and spatial resolution of mass reconstructions from space, it becomes possible to detect halos
the size of galaxy groups, as well as clusters over a wide range of redshifts -- thus tracing
their formation,} which is governed by the properties of dark matter and the nature of gravity. 
Space-based data also crosses the threshold to mapping even filamentary large-scale structure in three dimensions
\citep{massey07b}. 
Obtaining the detailed, 3D
distribution of mass will be particularly important near regions of interest like the Bullet cluster
\citep{clowe06}, where the small differences between the location of mass and baryons in a small
patch of sky may yield the best possible
information about the properties of dark matter. In this and other astrophysical phenomena, knowledge of the
local mass environment and nearby large-scale structure is critical.

Overall, we conclude that ground-based weak lensing surveys can perform several tasks remarkably
well, with sufficiently small amount of systematic bias to easily justify the next generation of
dedicated  ground-based surveys. Two dimensional statistical analyses will be able to produce 
order-of-magnitude improvements in weak lensing constraints, using proven hardware technology and
software pipelines. On the other hand, a wide-field space-based imager would provide important
control over some systematic effects, and open up many new applications that are, at least
currently, unachievable from the ground. For several of the most exciting techniques that will
directly probe the nature of dark matter and dark energy, eventual space-based imaging is likely to
be essential.

\section{Acknowledgments}

This work is supported by the US Department of Energy under contract DE-FG02-04ER41316. It is based upon observations with the NASA/ESA
Hubble Space Telescope, obtained at the  Space Telescope Science Institute, which is operated by AURA Inc, under NASA contract  NAS 5-26555.
It is also based on data collected at the Subaru Telescope, operated by the  National Astronomical Observatory of Japan. It is our pleasure
to thank Alexie Leauthaud and Takashi Hamana for help with the catalogs and Alexandre Refregier, James Taylor and Alexis Finoguenov for
useful discussions. We also gratefully acknowledge the contributions of the entire COSMOS collaboration, consisting of more than 70
scientists worldwide. More information on the COSMOS survey is available from {\tt \url{http://www.astro.caltech.edu/$\sim$cosmos}}.


\end{document}